# Constraint Inequalities from Hilbert Space Geometry & Efficient Quantum Computation

Chinonso Onah

Arnold Sommerfeld Center & Faculty of Physics, Ludwig-Maximilians-Universität, Munich

## Introduction

Useful relations describing arbitrary parameters of given quantum systems can be derived from simple physical constraints imposed on the vectors in the corresponding Hilbert space. This is well known and it usually proceeds by partitioning the large dimensional Hilbert space into relevant sub spaces and relating points in the Hilbert space to the expectation values of physical observables. The aim of this note is quite modest. We describe the procedure and point out that this parallels the necessary considerations that make Quantum Simulation of quantum fields and interacting many body quantum systems on Noisy Intermediate Scale Quantum (NISQ) devices possible. We conclude by pointing out relevant parts of Quantum Computing where these ideas could be useful. This work proceeds in density matrix formalism and is a review of materials found in references. We enrich the literature by suggesting how to use these ideas to guide and improve parameterised quantum circuits.

## Formalism

Recall that a complex N × N matrix is a density matrix if it is

Hermitian ($\rho = \rho^\dagger$),

Positive and ($\rho \geq 0$), and         (1)

Normalised (Tr $\rho = 1$).

These complex matrices define a Hilbert space and in addition

det $\rho \geq 0$.         (2)

Straight lines, called Hilbert-Schmidt distance, exist in this space. It is given by $D_2^2(\rho, \rho') = \frac{1}{2}(\rho - \rho')^2$ for $\rho$ and $\rho'$ in the Hilbert space. This space is also convex since an arbitrary state can be written as a linear superposition or a statistical mixture of pure states. With convexity in mind, exploiting the constraints in (1) and (2) leads to useful characterization of parameters of an arbitrary density matrix and the underlying geometry of the space.

## Spin ½ Systems

Recall that the number of dof in an **n** X **n** Hermitian matrix whose first r columns are linearly independent, and the remaining **n - r** columns are linear combinations of the first r is  r(2n – r)         (3)

This implies the set of states in spin-1/2 systems have 2 strata of ranks 1 and 2 corresponding to subsystems of dimensions 2 & 3 respectively.

An arbitrary 2 X 2 Hermitian matrix is given by

$$\rho = \begin{bmatrix} a & \alpha - i\beta \\ \alpha + i\beta & b \end{bmatrix} \qquad (4)$$

To define a density matrix we require the following constraints w.l.o.g

$0 \leq a \leq 1; \ 0 \leq b \leq 1; \ a + b = 1; \ \alpha^2 + \beta^2 \leq ab$

Which restricting to a 3-D hyperplane we have

$$(a - \tfrac{1}{2})^2 + \alpha^2 + \beta^2 \leq \tfrac{1}{4} \qquad (5)$$

$\Rightarrow (\alpha, \beta, a)$ lie inside or on the ball centered on $(0,0,1/2)$

The points are in (1,1) correspondence with unit rays in $\mathbb{C}^2$ and we have a mapping from $S^3 \subset \mathbb{C}^2 \to S^2$ in $\mathbb{R}^3$ (a Hopf fibration).

If we consider the so called Cartesian or Mixture Coordinate we set

$$\rho = \tfrac{1}{N}\mathbb{I} + \sum_{i=1}^{N^2-1} \tau_i \sigma_i \qquad (6)$$

Where $\sigma_i \sigma_j = \tfrac{2}{N}\delta_{ij} + d_{ijk}\sigma_k + i f_{ijk}\sigma_k$.

$$\rho^2 = \rho \implies \begin{cases} \tau^2 = \tfrac{N-1}{2N} \\ \tfrac{N-2}{N}\tau_i \end{cases} \qquad (7)$$

$\Rightarrow$ Pure states get confined to a $(N^2-2)$-Dimensional outsphere and in addition they form a well defined subset of the surface of the outsphere for $N > 2$. This defines a **Bloch Ball** with mixed states residing inside.

The mixed states have rank 2 $\Rightarrow$ Sum of 2 states is given by drawing a line in $\rho$ space.

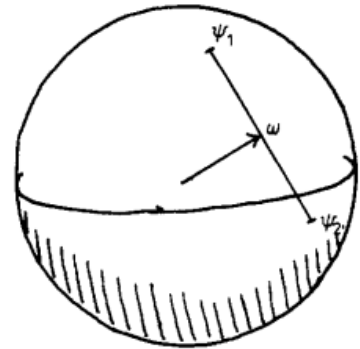

Considering $\vec{J} = (Jx, Jy, Jz) =$

$\tfrac{1}{2}\left( \begin{pmatrix} 1 & 0 \\ 0 & 1 \end{pmatrix}, \begin{pmatrix} 0 & -i \\ i & 0 \end{pmatrix}, \begin{pmatrix} 1 & 0 \\ 0 & -1 \end{pmatrix} \right)$ leads to

$\langle J_x \rangle = \alpha; \ \langle J_y \rangle = \beta$ and $\langle J_z \rangle = a - 1/2$

$\langle J_x \rangle^2 + \langle J_y \rangle^2 + \langle J_z \rangle^2 = (a - \tfrac{1}{2})^2 + \alpha^2 + \beta^2 = \tfrac{1}{4}$

## Spin -1 Systems

The convex set of states has 3 strata with ranks 1,2,3 resp. dim 4,7,8. Matrices of rank 3 form the convex region of $R^8$. Pure states of rank 1 form 4-dimensional sub region.

$$Let\ \rho = \begin{pmatrix} a & \bar{h} & g \\ h & b & \bar{f} \\ \bar{g} & f & c \end{pmatrix}; a + b + c = 1 \qquad (8)$$

**SPIN-1 Constraints: CASE I**

$a \geq 0; b \geq 0; c \geq 0; |f|^2 \leq bc; |g|^2 \leq ca; |h|^2 \leq ab;$ and

$\det\rho = abc + 2\text{Re}(fgh) - (a|f|^2 + b|g|^2 + c|h|^2) \geq 0 \qquad (9)$

$a + b + c = 1; a \geq 0; b \geq 0; c \geq 0;$ lies in positive octant. Each point in $\triangle abc$ corresponds to an allowed set of values (a,b,c) & over each point there is a convex 6D region ($f_r, f_i, g_r, g_i, h_r, h_i$).

$|f|^2 \leq bc; |g|^2 \leq ca; |h|^2 \leq ab \Rightarrow$ the following 2-D extremal discs at the vertices $|f|^2 \leq bc$, g=h=0; $|g|^2 \leq ca$, f=h=0; and $|h|^2 \leq ab$, f=g=0

**SPIN-1 Constraints: CASE II**

Another class constraints can be defined. For example, a more direct way to determine various sub regions is to set f=$\sqrt{bc}F$; $g = \sqrt{ca}G$; & $h = \sqrt{ab}H$

$\Rightarrow |F| \leq 1; |G| \leq 1; |H| \leq 1$

Then let F=$e^{i\mathcal{X}_F}$ & (9) leads to $|G|^2 + |H|^2 - 2\text{Re}(e^{i\mathcal{X}_F}GH) \leq 0 \Rightarrow$

$|Ge^{i\mathcal{X}_F} - \bar{H}|2 \leq 0$ ie $Ge^{i\mathcal{X}_F} = \bar{H} \qquad (10)$

I.e. (G,H) 4-space is restricted to a 2-plane through the origin

Then $|G| \leq 1$ & $|H| \leq 1$ gives the disc as above. We can also derive the phase constraint $\mathcal{X}_F + \mathcal{X}_G + \mathcal{X}_H = 0$ from (10)

## Constraining Qudits with Generalized Gell-Mann Matrix Basis

The generalized Gell-Mann matrices (GGM) are higher–dimensional generalization of the Pauli matrices and hence find application in study and analysis of multi-level quantum systems—called qudits. Three level systems correspond to 3x3 matrices. In this space is spanned by a set of the well-known Gell-Mann matrices and the corresponding quantum systems are called qutrits.

The GGM split into three species and lend themselves to compact operator notation; then the density matrices follow by simply writing the operators in the standard basis [5]

    i.    d(d−1)/2 symmetric GGM

$$\Lambda_s^{jk} = |j\rangle\langle k| + |k\rangle\langle j|, \quad 1 \leq j < k \leq d, \qquad 1 \leq j < k \leq d \qquad (11)$$

ii.  d(d−1)/2 antisymmetric GGM

$$\Lambda_a^{jk} = -i\,|j\rangle\langle k| + i\,|k\rangle\langle j| \qquad 1 \le j < k \le d \qquad (12)$$

iii.  (d − 1) diagonal GGM

$$\Lambda^l = \sqrt{\frac{2}{l(l+1)}}\left(\sum_{j=1}^{l} |j\rangle\langle j| - l\,|l+1\rangle\langle l+1|\right) \qquad 1 \le l \le d-1 \qquad (13)$$

The correspond to $d^2-1$ matrices.

Why are these GGMs important? They furnish a representation of the Hilbert space of any d-level quantum system and thus, thanks to GGMs, derivation of constraint inequalities could be easily carried out for arbitrary quantum systems.

For d=3 we have:

i.  3 symmetric Gell-Mann matrices

$$\lambda_s^{12} = \begin{pmatrix} 0 & 1 & 0 \\ 1 & 0 & 0 \\ 0 & 0 & 0 \end{pmatrix},\quad \lambda_s^{13} = \begin{pmatrix} 0 & 0 & 1 \\ 0 & 0 & 0 \\ 1 & 0 & 0 \end{pmatrix},\quad \lambda_s^{23} = \begin{pmatrix} 0 & 0 & 0 \\ 0 & 0 & 1 \\ 0 & 1 & 0 \end{pmatrix}, \qquad (14)$$

ii.  3 anti-symmetric Gell-Mann matrices

$$\lambda_a^{12} = \begin{pmatrix} 0 & -i & 0 \\ i & 0 & 0 \\ 0 & 0 & 0 \end{pmatrix},\quad \lambda_a^{13} = \begin{pmatrix} 0 & 0 & -i \\ 0 & 0 & 0 \\ i & 0 & 0 \end{pmatrix},\quad \lambda_a^{23} = \begin{pmatrix} 0 & 0 & 0 \\ 0 & 0 & -i \\ 0 & i & 0 \end{pmatrix}, \qquad (15)$$

iii.  2 diagonal Gell-Mann matrices

$$\lambda^1 = \begin{pmatrix} 1 & 0 & 0 \\ 0 & -1 & 0 \\ 0 & 0 & 0 \end{pmatrix},\quad \lambda^2 = \frac{1}{\sqrt{3}}\begin{pmatrix} 1 & 0 & 0 \\ 0 & 1 & 0 \\ 0 & 0 & -2 \end{pmatrix}. \qquad (16)$$

Each matrix in equations (14) to (16) can be derived from (8) by fixing the parameters a, b,c, d, e, f, g, h implying that the set of constraints can be as general or specific as necessary. These constraints can in turn be employed to analyze properties of the quantum hardware through quantities such as information entropy and evolution of quantum trajectory.

## The Parallel with Quantum Simulation

Consider the following. In 0+1-D QED, the dynamical degrees of freedom are two fermion sites and two flux links. However, there are two quantum states per fermion site and three per fermion link leading to 36 states. That would require 6 qubits since a qubit can only represent 2 quantum states. The associated charges are Q=0+1 and -1 and, thanks to gauss' law, divides the space into three different sectors.

Well known discrete spacetime symmetries can be further utilized to decompose each charge sector. Take the Q=0 for example. It has 5 states and the parity operator can be utilized to split the subsector into P=+1 sector which has 3 states and an be described with 2 qubits and P=-1 sector which has 2 states and require only 1 qubit. The conclusion is immediate: instead of 6 qubits, we need only 2 to describe QED 0+1 dimension.

The case for 1+1 D is even more striking. The dynamical dof are 4 fermion sites and 4 links leading to 1296 states (naively) leading to a 12 qubit requirement. Utilization of available constraints reduces the requirements to merely 3 qubits! This is certainly good news for NISQ era and we refer to section II and Appendix G of ref [3] for a detailed analysis. The parallel we seek to emphasize is the use of physical constraints to reduce the huge dimensionality of an otherwise large problem into manageable sizes. And to this end general form of constraints can be a useful tool.

**Possible Applications**

We suggest that it is possible to use these ideas to train quantum circuits parameterized by multiple parameters by explicitly using the constraint relations derived from physical principles in optimizations. In Quantum Deep Neural Network for example, the Quantum agent traverses a large Hilbert space to find optimal solution and frequently strays from the optimal solution after learning ( see [6] and [7] for an example). Thus, constraining the search to only physically relevant part of the Hilbert space could improve time of convergence and the overall quality of solutions. Similarly, deriving arbitrary parameterization will also be relevant in computing gradients during the learning phase.

On the other hand, in many body systems, the associated Hilbert spaces are always exponential in size with over 99% of it being unphysical in certain cases. Finding clever restrictions to only physical relevant spaces is crucial for meaningful quantum computing in NISQ era. Experiments to determine if explicit use of constraint derived Hilbert space geometry leads to quantitative advantage in parameterized quantum circuits is ongoing. This has huge implications for Quantum Simulation, Quantum Circuit Compilation and Quantum optimization.

*Acknowledgements*

Obinna Abah, Obinna Uzoh, Deepak Aryal and A. Al-Eryani are acknowledged for helpful discussions and feedbacks.

*References*

(1) JE Harriman (1978)  [Phys. Rev. A 17, 1249 (1978)](#)

(2) F J Bloore (1976). *J. Phys. A: Math. Gen.* **9** 2059

(3) Klco, N. and Dumitrescu, E. F. and McCaskey, A. J. and Morris, T. D. and Pooser, R. C. and Sanz, M. and Solano, E. and Lougovski, P. and Savage, M. J. *Phys.Rev.A* 98 (2018) 3, 032331.

(4) Bengtsson, I., & Zyczkowski, K. (2006). *Geometry of Quantum States: An Introduction to Quantum Entanglement*. Cambridge: Cambridge University

(5) R. A. Bertlmann and P. Krammer (2008). Bloch vectors for qudits. arxiv.org/0806.1174.

(6) Moro, L., Paris, M.G.A., Restelli, M. *et al*. Quantum compiling by deep reinforcement learning. *Commun Phys* **4**, 178 (2021).

(7) Y. Kwak et al (2021). Introduction to Quantum Reinforcement Learning: Theory and PennyLane-based Implementation. arxiv.org/2108.06849